\documentclass[twocolumn,showpacs,preprintnumbers,amsmath,amssymb,prb]{revtex4}

\usepackage{graphicx}
\usepackage{dcolumn}
\usepackage{bm}
\usepackage{tabularx}
\usepackage{amsmath}

\begin{document}

\title{Nodeless superconductivity in Ca$_3$Ir$_4$Sn$_{13}$: evidence from quasiparticle heat transport}

\author{S. Y. Zhou, H. Zhang, X. C. Hong, B. Y. Pan, X. Qiu, W. N. Dong, X. L. Li, and S. Y. Li$^*$}

\affiliation{State Key Laboratory of Surface Physics, Department of
Physics, and Laboratory of Advanced Materials, Fudan University,
Shanghai 200433, P. R. China}

\date{\today}

\begin{abstract}
We report resistivity $\rho$ and thermal conductivity $\kappa$
measurements on Ca$_3$Ir$_4$Sn$_{13}$ single crystals, in which
superconductivity with $T_c \approx 7$ K was claimed to coexist with
ferromagnetic spin-fluctuations. Among three crystals, only one
crystal shows a small hump in resistivity near 20 K, which was
previously attributed to the ferromagnetic spin-fluctuations. Other
two crystals show the $\rho \sim T^2$ Fermi-liquid behavior at low
temperature. For both single crystals with and without the
resistivity anomaly, the residual linear term $\kappa_0/T$ is
negligible in zero magnetic field. In low fields, $\kappa_0(H)/T$
shows a slow field dependence. These results demonstrate that the
superconducting gap of Ca$_3$Ir$_4$Sn$_{13}$ is nodeless, thus rule
out nodal gap caused by ferromagnetic spin-fluctuations.

\end{abstract}


\maketitle

\section{Introduction}

The interplay between magnetism and superconductivity has been a
central issue in unconventional superconductors. While the static
magnetism is generally believed to compete with superconductivity,
the dynamic magnetism could be the source of electron pairing.
\cite{MNorman} For example, the antiferromagnetic (AF)
spin-fluctuations are considered as the pairing glue in high-$T_c$
cuprates, iron-based superconductors, and many heavy-fermion
superconductors. \cite{MNorman} On another side, the ferromagnetic
(FM) spin-fluctuations could be the origin of the superconductivity
in Sr$_2$RuO$_4$, heavy-fermion superconductors UGe$_2$ and URhGe.
\cite{MNorman} These AF and FM spin-fluctuations usually result in
superconducting gaps with nodes, such as the $d$-wave gap in
cuprates and CeCoIn$_5$, \cite{CCTsuei,KAn} and the $p$-wave gap in
Sr$_2$RuO$_4$, \cite{Mackenzie} but in some cases, such as the
multiband iron-based superconductors, the AF spin-fluctuations may
give $s_{\pm}$-wave gap without nodes. \cite{PJHirschfeld}

Ca$_3$Ir$_4$Sn$_{13}$ is a cubic transition metal compound, in which
superconductivity with $T_c \approx$ 7 K was found thirty years ago.
\cite{Espinosa} Very few studies have been done on this compound
since its discovery. Until recently, detailed resistivity,
susceptibility, and specific heat measurements suggested that the
superconductivity coexists with the FM spin-fluctuations in
Ca$_3$Ir$_4$Sn$_{13}$. \cite{JinhuYang} A peak-like anomaly near 30
K in resistivity was observed. Below the anomaly, the
non-Fermi-liquid behavior of resistivity in zero field has been
attributed to the FM spin-fluctuations, and upon applying magnetic
field, Fermi liquid behavior is recovered. \cite{JinhuYang}

Since the FM spin-fluctuations may cause a nodal superconducting
state in Ca$_3$Ir$_4$Sn$_{13}$, it will be interesting to probe its
superconducting gap structure. Ultra-low-temperature thermal
conductivity measurement is such a bulk technique.
\cite{Shakeripour} The existence of a finite residual linear term
$\kappa_0/T$ in zero magnetic field is an evidence for gap nodes.
The field dependence of $\kappa_0/T$ may further give support for a
nodal superconducting state, and provide informations on the gap
anisotropy, or multiple gaps. \cite{RobHill}

In this paper, we measure the resistivity and thermal conductivity
of Ca$_3$Ir$_4$Sn$_{13}$ single crystals. We find that while one
crystal shows a small hump in resistivity near 20 K, other two
crystals show $\rho \sim T^2$ Fermi-liquid behavior in zero field.
For both single crystals with and without the resistivity anomaly,
the absence of $\kappa_0/T$ in zero field and the slow field
dependence of $\kappa_0(H)/T$ in low fields clearly demonstrate
nodeless superconductivity in Ca$_3$Ir$_4$Sn$_{13}$.

\section{Experiment}

Single crystals of Ca$_3$Ir$_4$Sn$_{13}$ were grown by flux method,
as previously described in Ref. 6. The excessive Sn flux was etched
in concentrated hydrochloric acid (HCl). The obtained single
crystals have typical size of a few mm$^3$. We chose three single
crystals with large flat surface, which was identified as (110)
plane by X-ray diffraction measurements. Then the single crystals
were polished and cut to a rectangular shape of typical dimensions
2.5 $\times$ 1.0 mm$^2$ in the (110) plane, and 0.2 mm in thickness.
The dc magnetic susceptibility was measured in $H$ = 20 Oe both
parallel and perpendicular to the (110) plane, using a SQUID (MPMS,
Quantum Design). Four silver wires were attached on the sample with
silver paint, which were used for both resistivity and thermal
conductivity measurements, with electrical and heat currents in the
(110) plane. The contacts are metallic with typical resistance 50
m$\Omega$ at 2 K. The thermal conductivity was measured in a
dilution refrigerator, using a standard four-wire steady-state
method with two RuO$_2$ chip thermometers, calibrated {\it in situ}
against a reference RuO$_2$ thermometer. Magnetic fields were
applied perpendicular to the (110) plane. To ensure a homogeneous
field distribution in the sample, all fields were applied at
temperature above $T_c$.

\section{Results and Discussion}

Figure 1(a) presents the typical dc magnetic susceptibility of
Ca$_3$Ir$_4$Sn$_{13}$ single crystals, measured in $H$ = 20 Oe
parallel and perpendicular to the (110) plane, with zero-field
cooled. The transition temperature $T_c \approx$ 6.9 K is determined
from the onset of diamagnetic transition. In field perpendicular to
the (110) plane, the shielding volume fractions exceeds -1 at 2 K,
indicating bulk superconductivity of the sample. Fig. 1(b) plots the
resistivities of samples S1, S2, and S3 in zero magnetic field. For
all three samples, the zero-resistivity $T_c \approx$ 7.25 K is
slightly higher than that obtained from the magnetization
measurements in Fig. 1(a), and the 10-90\% resistive transition
widths are less than 0.2 K.

\begin{figure}
\includegraphics[clip,width=5.0cm]{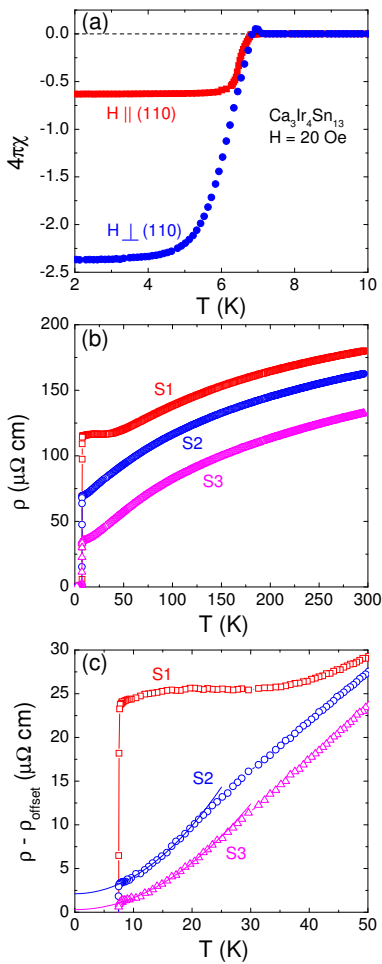}
\caption{(Color online) (a) The typical dc magnetic susceptibility
of Ca$_3$Ir$_4$Sn$_{13}$ single crystals measured in $H$ = 20 Oe
both parallel and perpendicular to the (110) plane, with zero-field
cooled. (b) Resistivity of three samples in zero magnetic field. (c)
Low-temperature data of the resistivities in (b). The data sets are
offset for clarity. Both samples S2 and S3 show $\rho \sim T^2$
Fermi-liquid behavior at low temperature, as indicated by the solid
lines.}
\end{figure}

\begin{figure}
\includegraphics[clip,width=5.3cm]{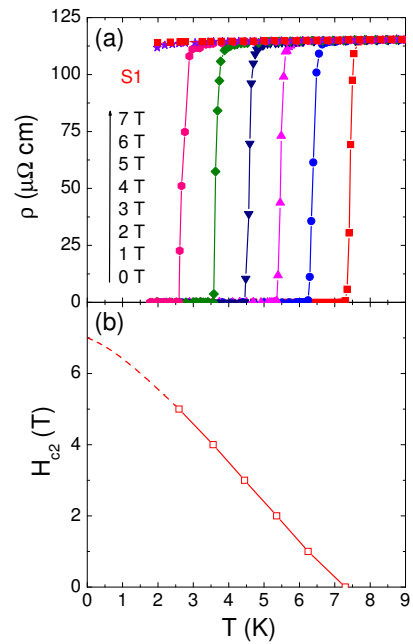}
\caption{(Color online) (a) Resistivity of Ca$_3$Ir$_4$Sn$_{13}$
single crystal S1 in $H$ = 0, 1, 2, 3, 4, 5, 6, and 7 T. The
normal-state $\rho$(7T) curve is quite flat, extrapolating to
residual resistivity $\rho_0$(7T) $\approx$ 114 $\mu\Omega$ cm. (b)
Temperature dependence of the upper critical field $H_{c2}$, defined
by $\rho$ = 0. The dashed line is a guide to the eye, which points
to $H_{c2}(0) \approx$ 7 T.}
\end{figure}

From Fig. 1(b), only the resistivity of S1 shows an anomaly at low
temperature, while S2 and S3 show good metallic behavior. Fig. 1(c)
plots the resistivity data blow 50 K. All data sets are offset for
clarity. One can see that the small hump near 20 K of S1 is much
less pronounced than the peak near 30 K in Ref. 7. Interestingly,
for S2 and S3, the low-temperature data obey the Fermi-liquid
behavior $\rho \sim T^2$, as indicated by the solid lines. We note
that the residual resistivity $\rho_0$ of S1 is the highest among
all three samples. If lower $\rho_0$ indicates higher sample
quality, the resistivity anomaly may be not an intrinsic property of
Ca$_3$Ir$_4$Sn$_{13}$. However, we are not sure about this at
current stage.

Nevertheless, it is necessary to check the superconducting gap
structure of both Ca$_3$Ir$_4$Sn$_{13}$ samples with and without the
resistivity anomaly. The upper critical field $H_{c2}$ needs to be
determined first. Fig. 2(a) shows the resistivity of
Ca$_3$Ir$_4$Sn$_{13}$ single crystal S1 in magnetic field up to $H$
= 7 T. For $H$ = 7 T, no superconducting transition is observed down
to 2 K. The normal-state $\rho$(7T) curve is quite flat,
extrapolating to a residual resistivity $\rho_0$(7T) = 114.0
$\mu\Omega$ cm. In Fig. 2(b), we plot the temperature dependence of
the upper critical field $H_{c2}$, defined by $\rho$ = 0 from the
curves in Fig. 2(a). A rough estimation gives $H_{c2}(0) \approx$ 7
T. Similar measurements on S2 give the same $H_{c2}(0)$, and
$\rho_0$(7T) = 79.6 $\mu\Omega$ cm.

Figure 3(a) shows the temperature dependence of the thermal
conductivity for samples S1 and S2 in zero magnetic fields, plotted
as $\kappa/T$ vs $T$. The measured thermal conductivity is the sum
of two contributions, respectively from electrons and phonons, so
that $\kappa = \kappa_e + \kappa_p$. In order to obtain the residual
linear term $\kappa_0/T$ contributed by electrons, we extrapolate
$\kappa/T$ to $T = 0$. Usually, this can be done by fitting the data
to $\kappa/T = a + bT^{\alpha-1}$ at low temperature, where $a
\equiv \kappa_0/T$. \cite{Sutherland,SYLi1} The power $\alpha$ of
the second term contributed by phonons is typically between 2 and 3,
due to specular reflections of phonons at the boundary.
\cite{Sutherland,SYLi1} From Fig. 3(a), the curves below 250 mK can
be well fitted by $\kappa/T = a + bT^{\alpha-1}$ with $\alpha$ =
2.41 and 2.27 for S1 and S2, respectively. Previously $\alpha
\approx$ 2.2 has been observed in the $s$-wave superconductor
Cu$_{0.06}$TiSe$_2$, \cite{SYLi2} and recently $\alpha \approx$ 2
was found in some iron-based superconductors such as
BaFe$_{1.9}$Ni$_{0.1}$As$_2$ and KFe$_2$As$_2$. \cite{LDing,JKDong}
Below we will only focus on $\kappa_0/T$.

\begin{figure}
\includegraphics[clip,width=6.6cm]{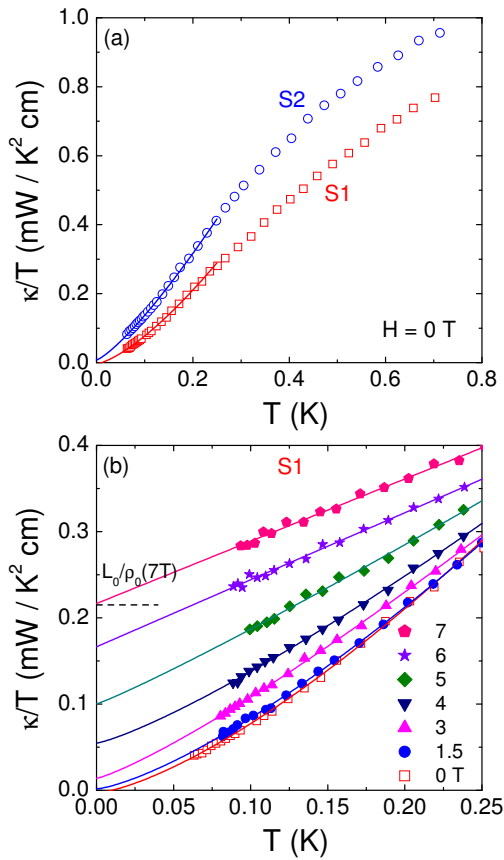}
\caption{(Color online) (a) Low-temperature thermal conductivity of
Ca$_3$Ir$_4$Sn$_{13}$ single crystals S1 and S2 in zero magnetic
field. The solid lines are $\kappa/T = a + bT^{\alpha-1}$ fits below
250 mK, respectively. (b) Thermal conductivity of S1 in magnetic
fields. The dash line is the normal-state Wiedemann-Franz law
expectation $L_0$/$\rho_0$(7T), with $L_0$ the Lorenz number 2.45
$\times$ 10$^{-8}$ W $\Omega$ K$^{-2}$ and $\rho_0$(7T) = 114.0
$\mu\Omega$ cm.}
\end{figure}

In zero field, the fittings give $\kappa_0/T$ = -3 $\pm$ 3 and 8
$\pm$ 5 $\mu$W K$^{-2}$ cm$^{-1}$ for S1 and S2, respectively. Note
that our experimental error bar is about 5 $\mu$W K$^{-2}$
cm$^{-1}$. For $s$-wave nodeless superconductors, there are no
fermionic quasiparticles to conduct heat as $T \to 0$, since all
electrons become Cooper pairs. Therefore there is no residual linear
term of $\kappa_0/T$, as seen in V$_3$Si. \cite{Sutherland} However,
for unconventional superconductors with nodes in the superconducting
gap, the nodal quasiparticles will contribute a finite $\kappa_0/T$
in zero field. \cite{Shakeripour} For example, $\kappa_0/T$ = 1.41
mW K$^{-2}$ cm$^{-1}$ for the overdoped cuprate
Tl$_2$Ba$_2$CuO$_{6+\delta}$ (Tl-2201), a $d$-wave superconductor
with $T_c$ = 15 K. \cite{Proust} For the $p$-wave superconductor
Sr$_2$RuO$_4$, $\kappa_0/T$ = 17 mW K$^{-2}$ cm$^{-1}$.
\cite{MSuzuki} Therefore, the negligible $\kappa_0/T$ of both S1 and
S2 samples strongly suggests that the superconducting gap of
Ca$_3$Ir$_4$Sn$_{13}$ is nodeless.

Figure 3(b) plots the thermal conductivity of S1 in magnetic fields.
All the curves are also fitted by $\kappa/T = a + bT^{\alpha-1}$. In
$H_{c2}$ = 7 T, $\kappa_0/T$ = 0.216 $\pm$ 0.003 mW K$^{-2}$
cm$^{-1}$ was obtained from the fitting. This value meets the
Wiedemann-Franz law expectation $L_0/\rho_0$(7T) = 0.215 mW K$^{-2}$
cm$^{-1}$ nicely, with $L_0$ the Lorenz number 2.45 $\times$
10$^{-8}$ W$\Omega$K$^{-2}$ and $\rho_0$(7T) = 114.0 $\mu\Omega$ cm.
The verification of Wiedemann-Franz law in the normal state shows
the reliability of our thermal conductivity measurements.

As seen in Fig. 3(b), $\kappa_0/T$ of S1 gradually increases with
increasing field. In Fig. 4, we plot the normalized $\kappa_0(H)/T$
as a function of $H/H_{c2}$ for S1. The $\kappa_0/T$ of S2 shows
similar field dependence, which is not shown here. For comparison,
the data of clean $s$-wave superconductor Nb, \cite{Lowell} the
dirty $s$-wave superconducting alloy InBi, \cite{Willis} the
multi-band $s$-wave superconductor NbSe$_2$, \cite{Boaknin1} and an
overdoped sample of the $d$-wave superconductor Tl-2201
\cite{Proust} are also plotted. For a clean type-II $s$-wave
superconductor with a single gap, $\kappa$ should grow exponentially
with field (above $H_{c1}$), as is indeed observed in Nb.
\cite{Lowell} For InBi, the curve is exponential at low $H$,
crossing over to a roughly linear behavior closer to $H_{c2}$ as
expected for $s$-wave superconductors in the dirty limit.
\cite{Caroli}

\begin{figure}
\includegraphics[clip,width=6.6cm]{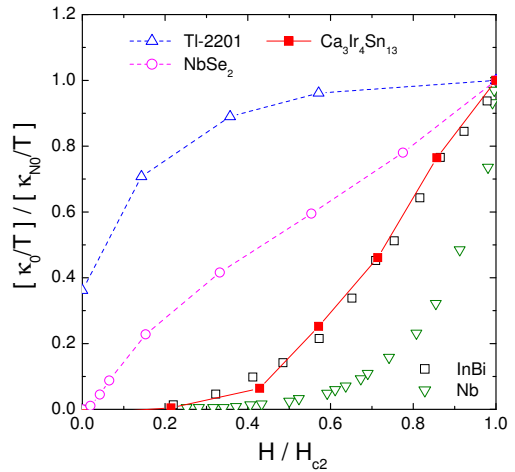}
\caption{(Color online) Normalized residual linear term $\kappa_0/T$
of Ca$_3$Ir$_4$Sn$_{13}$ plotted as a function of $H/H_{c2}$. For
comparison, similar data are shown for the clean $s$-wave
superconductor Nb, \cite{Lowell} the dirty $s$-wave superconducting
alloy InBi, \cite{Willis} the multi-band $s$-wave superconductor
NbSe$_2$, \cite{Boaknin1} and an overdoped sample of the $d$-wave
superconductor Tl-2201. \cite{Proust}}
\end{figure}

The normalized $\kappa_0(H)/T$ of Ca$_3$Ir$_4$Sn$_{13}$ clearly
mimics that of the dirty $s$-wave superconductor InBi. However,
previously Yang {\it et al.} estimated the superconducting coherence
length $\xi_0 \sim$ 79 \AA\ and electronic mean free path $l \sim$
811 \AA, which implies that Ca$_3$Ir$_4$Sn$_{13}$ is an intrinsic
clean-limit superconductor ($\xi_0 \ll l$). \cite{JinhuYang} Since
the $\rho_0$ of our sample S1 is about twice that of Yang {\it et
al.}'s, we estimate the mean free path $l \sim$ 400 \AA\ for S1. The
$H_{c2}(0)$ of S1 is the same as in Ref. 7, which gives the same
$\xi_0 \sim$ 79 \AA\ through the relation $H_{c2}(0)$ =
$\Phi_0/2\pi\xi_0{^2}$. Therefore sample S1 is still a clean
superconductor, and the relatively fast increase of $\kappa_0(H)/T$
near $H_{c2}$ should have different origin from the dirty $s$-wave
superconductor InBi.

Band structure calculation shows that there are six bands in
Ca$_3$Ir$_4$Sn$_{13}$ which cross the Fermi level. \cite{SKGoh} Each
sheet of the Fermi surface was found to be three dimensional, with
rather complex shape. \cite{SKGoh} In this context, we interpret the
$\kappa_0(H)/T$ behavior of Ca$_3$Ir$_4$Sn$_{13}$ may result from
multiple gaps or gap anisotropy.

For the multi-band $s$-wave superconductor NbSe$_2$, $\kappa_0(H)/T$
increases rapidly in both low field and near $H_{c2}$.
\cite{Boaknin1} Similar behavior has been observed in
$L$Ni$_2$B$_2$C ($L$ = Y, Lu) with multiple and anisotropic gaps.
\cite{Boaknin2,TBaba} In both NbSe$_2$ and $L$Ni$_2$B$_2$C, applying
a field rapidly delocalizes quasiparticle states confined within the
vortices associated with the smaller gap band, while those states
associated with the larger gap band delocalize more slowly. For
NbSe$_2$, the ratio between larger and smaller gaps is approximately
3, \cite{TYokoya} and for YNi$_2$B$_2$C, the ratio is about 2.1.
\cite{TBaba} Recently, Bang has calculated the field dependence of
$\kappa_0(H)/T$ for different gap ratios, to explain the thermal
conductivity data of multi-gap iron-based superconductor
Ba(Fe$_{1-x}$Co$_x$)$_2$As$_2$. \cite{YBang} It is possible that in
Ca$_3$Ir$_4$Sn$_{13}$, the gaps in the six Fermi surfaces may have
different magnitudes, or in some Fermi surface the gap is
anisotropic. If this is the case, according to Bang's calculation,
\cite{YBang} the gap ratio in Ca$_3$Ir$_4$Sn$_{13}$ should be around
1.4 or so. This interpretation needs to be checked by momentum
dependent measurements of the superconducting gap, such as
angleresolved photoemission spectroscopy (ARPES) experiments.

\section{Summary}

In summary, we report the resistivity and thermal conductivity
measurements of Ca$_3$Ir$_4$Sn$_{13}$ single crystals. Among three
crystals, only one sample shows resistivity anomaly near 20 K, and
other two samples display Fermi-liquid behavior $\rho \sim T^2$ at
low temperature. This suggests that the resistivity anomaly may be
not intrinsic in Ca$_3$Ir$_4$Sn$_{13}$. Thermal conductivity results
clearly demonstrate that the superconducting gap is nodeless in both
crystals with and without the resistivity anomaly. This implies that
the FM spin-fluctuations, if exist, may be irrelevant to the
superconductivity in Ca$_3$Ir$_4$Sn$_{13}$, and the conventional
electron-phonon interaction should be responsible for the electron
pairing. The $\kappa_0(H)/T$ shows a relatively fast increase near
$H_{c2}$. Since Ca$_3$Ir$_4$Sn$_{13}$ is not in the dirty limit, but
rather has multiple Fermi surfaces with complex shape, we interpret
that the behavior of $\kappa_0(H)/T$ may result from gap anisotropy,
or multiple isotropic gaps with different magnitudes.

\begin{center}
{\bf ACKNOWLEDGEMENTS}
\end{center}

This work is supported by the Natural Science Foundation of China,
the Ministry of Science and Technology of China (National Basic
Research Program No: 2009CB929203 and 2012CB821402), Program for
Professor of Special Appointment (Eastern Scholar) at Shanghai
Institutions of Higher Learning.\\

$^*$ E-mail: shiyan$\_$li@fudan.edu.cn


\begin{thebibliography}{99}

\bibitem{MNorman} M. R. Norman, Science {\bf 332}, 196 (2011), and references therein.
\bibitem{CCTsuei} C. C. Tsuei and J. R. Kirtley, Rev. Mod. Phys. {\bf 72}, 969 (2000).
\bibitem{KAn} K. An, T. Sakakibara, R. Settai, Y. Onuki, M. Hiragi, M. Ichioka, and K. Machida, Phys. Rev. Lett. {\bf 104}, 037002 (2010).
\bibitem{Mackenzie} A. P. Mackenzie and Y. Maeno, Rev. Mod. Phys. {\bf 75}, 657 (2003).
\bibitem{PJHirschfeld} P. J. Hirschfeld, M. M. Korshunov, and I. I. Mazin, Rep. Prog. Phys. {\bf 74}, 124508 (2011).
\bibitem{Espinosa} G. P. Espinosa, Mater. Res. Bull. {\bf 15}, 791 (1980). G. P. Espinosa, A. S. Copper, and H. Barz, Mater. Res. Bull. {\bf 17}, 963 (1982).
\bibitem{JinhuYang} Jinhu Yang, Bin Chen, Chishiro Michioka, and Kazuyoshi Yoshimura, J. Phys. Soc. Jpn. {\bf 19}, 113705 (2010).
\bibitem{Shakeripour} H. Shakeripour, C. Petrovic, and L. Taillefer, New J. Phys. {\bf 11}, 055065 (2009).
\bibitem{RobHill} R. W. Hill, Shiyan Li, M. B. Maple, and Louis Taillefer, Phys. Rev. Lett. {\bf 101}, 237005 (2008).
\bibitem{Sutherland} M. Sutherland, D. G. Hawthorn, R. W. Hill, F. Ronning, S. Wakimoto, H. Zhang, C. Proust, E. Boaknin, C. Lupien, and Louis Taillefer, Phys. Rev. B {\bf 67}, 174520 (2003).
\bibitem{SYLi1} S. Y. Li, J.-B. Bonnemaison, A. Payeur, P. Fournier, C. H. Wang, X. H. Chen, and L. Taillefer, Phys. Rev. B {\bf 77}, 134501 (2008).
\bibitem{SYLi2} S. Y. Li, G. Wu, X. H. Chen, and Louis Taillefer, Phys. Rev. Lett. {\bf 99}, 107001 (2007).
\bibitem{LDing} L. Ding, J. K. Dong, S. Y. Zhou, T. Y. Guan, X. Qiu, C. Zhang, L. J. Li, X. Lin, G. H. Cao, Z. A. Xu and S. Y. Li, New J. Phys. {\bf 11}, 093018 (2009).
\bibitem{JKDong} J. K. Dong, S. Y. Zhou, T. Y. Guan, H. Zhang, Y. F. Dai, X. Qiu, X. F. Wang, Y. He, X. H. Chen, and S. Y. Li, Phys. Rev. Lett. {\bf 104}, 087005 (2010).
\bibitem{MSuzuki} M. Suzuki, M. A. Tanatar, N. Kikugawa, Z. Q. Mao, Y. Maeno, and T. Ishiguro, Phys. Rev. Lett. {\bf 88}, 227004 (2002).
\bibitem{Lowell} J. Lowell and J. B. Sousa, J. Low. Temp. Phys. {\bf 3}, 65 (1970).
\bibitem{Willis} J. O. Willis and D. M. Ginsberg, Phys. Rev. B {\bf 14}, 1916 (1976).
\bibitem{Boaknin1} E. Boaknin, M. A. Tanatar, J. Paglione, D. Hawthorn, F. Ronning, R. W. Hill, M. Sutherland, Louis Taillefer, J. Sonier, S. M. Hayden, and J. W. Brill, Phys. Rev. Lett. {\bf 90}, 117003 (2003).
\bibitem{Proust} C. Proust, E. Boaknin, R. W. Hill, Louis Taillefer, and A. P. Mackenzie, Phys. Rev. Lett. {\bf 89}, 147003 (2002).
\bibitem{Caroli} C. Caroli and M. Cyrot, Phys. Kondens. Mater. {\bf 4}, 285 (1965).
\bibitem{Boaknin2} E. Boaknin, R. W. Hill, C. Proust, C. Lupien, Louis Taillefer, and P. C. Canfield, Phys. Rev. Lett. {\bf 87}, 237001 (2001).
\bibitem{TBaba} T. Baba, T. Yokoya, S. Tsuda, T. Watanabe, M. Nohara, H. Takagi, T. Oguchi, and S. Shin, Phys. Rev. B {\bf 81}, 180509(R) (2010).
\bibitem{TYokoya} T. Yokoya, T. Kiss, A. Chainani, S. Shin, M. Nohara, H. Takagi, Science {\bf 294}, 2518 (2001).
\bibitem{SKGoh} S. K. Goh, L. E. Klintberg, P. L. Alireza, D. A. Tompsett, Jinhu Yang, Bin Chen, K. Yoshimura, and F. Malte Grosche, arXiv:1105.3941 (2011).
\bibitem{YBang} Yunkyu Bang, Phys. Rev. Lett. {\bf 104}, 217001 (2010).

\end{thebibliography}
\end{document}